\documentclass[usegraphicx,usenatbib,useAMS]{mn2e}

\usepackage{graphicx}
\usepackage{anysize}
\usepackage{subfigure}
\usepackage{amsmath}
\marginsize{2cm}{2cm}{1cm}{2cm} 

\newcommand{\kms}{km\,s$^{-1}$}

\newcommand{\lya}{Ly$\alpha$}
\newcommand{\cv}{C\,{\sc v}}
\newcommand{\civ}{C\,{\sc iv}}
\newcommand{\cii}{C\,{\sc ii}}
\newcommand{\ciii}{C\,{\sc iii}]}
\newcommand{\aiii}{Al\,{\sc iii}}
\newcommand{\siv}{Si\,{\sc iv}}
\newcommand{\nv}{N\,{\sc v}}
\newcommand{\hb}{H\,{\sc $\beta$}}

\newcommand{\mgii}{Mg\,{\sc ii}}
\newcommand{\lam}{$\lambda$}
\newcommand{\lala}{$\lambda$$\lambda$}
\newcommand{\aj}{Astron.~J. }%
\newcommand{\apj}{Astrophys.~J. }%
\newcommand{\apjl}{Astrophys. J. Lett. }%
\newcommand{\apjs}{Astrophys. J. Suppl. Ser. }%
\newcommand{\ao}{Appl. Opt. }%
\newcommand{\aap}{Astron. Astrophys. }%
\newcommand{\aaps}{Astron. Astrophys. Suppl. Ser. }%
\newcommand{\mnras}{Mon. Not. R. Astron. Soc. }%
\newcommand{\na}{New Astron. }%
\title[High-Velocity Emergent Broad Absorption]{Multi-Epoch Observations of Extremely High-Velocity Emergent Broad Absorption}

\author[J.A. Rogerson et al.]{
Jesse A. Rogerson,$^{1}$ \thanks{E-mail: rogerson@yorku.ca}
Patrick B. Hall,$^{1}$
Paola Rodr\'{i}guez Hidalgo,$^{1,2}$
\newauthor
Patrik Pirkola,$^{1}$
William N. Brandt,$^{3,4,5}$
Nur Filiz Ak$^{6}$\\
$^{1}$Department of Physics and Astronomy, York University, Toronto, ON M3J 1P3, Canada\\
$^{2}$Department of Physics and Astronomy, Humbolt State University, Arcata, CA 95521, USA\\
$^{3}$Department of Astronomy \&\ Astrophysics, 525 Davey Lab, The Pennsylvania State University, University Park, PA 16802, USA \\
$^{4}$Institute for Gravitation and the Cosmos, The Pennsylvania State University, University Park, PA 16802, USA \\
$^{5}$Department of Physics, 104 Davey Lab, The Pennsylvania State University, University Park, PA 16802, USA \\
$^{6}$Department of Astronomy and Space Sciences, Faculty of Sciences, Erciyes University, 38039 Kayseri, Turkey\\
}
\begin{document}
\maketitle

\begin{abstract}
We present the discovery of the highest velocity \civ\ broad absorption line
to date in the $z$=2.47 quasar SDSS J023011.28+005913.6, hereafter J0230.
In comparing the public DR7 and DR9 spectra of J0230, we discovered an emerging
broad absorption trough outflowing at $\sim$60,000~\kms, which we refer to as
trough A. In pursuing follow up observations of trough A, we discovered a
second emergent \civ\ broad absorption trough outflowing at $\sim$40,000~\kms,
namely trough B. In total, we collected seven spectral
epochs of J0230 that demonstrate emergent and rapidly ($\sim$10 days in the
rest-frame) varying broad absorption.
We investigate two possible scenarios that could cause these rapid changes:
bulk motion and ionization variability. Given our multi-epoch data, we
were able to rule out some simple models of bulk motion, but have proposed
two more realistic models to explain the variability of both troughs.
Trough A is likely an augmented `crossing disk' scenario with the absorber
moving at $10,000 < v~(\textrm{\kms}) < 18,000$. Trough B can be explained
by a flow-tube feature travelling
across the emitting region at $8,000 < v~(\textrm{\kms}) < 56,000$.
If ionization variability is the cause for the changes observed,
trough A's absorber has $n_e \geq 724 {\rm~cm}^{-3}$ and is at
$r_{equal} \ge 2.00 {\rm~kpc}$, or is at $r < 2.00 {\rm~kpc}$
with no constraint on the density; trough B's absorber either
has $n_e \geq 1540 {\rm~cm}^{-3}$ and is at
$r_{equal} \ge 1.37 {\rm~kpc}$, or is at $r < 1.37 {\rm~kpc}$
with no constraint on the density.

\end{abstract}


\section{Introduction}


At least 23\%\ of quasars exhibit blueshifted Broad Absorption Line (BAL)
troughs at ultraviolet wavelengths (see discussions in
\citealt{RH11} and \citealt{AH11}), and the fraction increases if narrower
(500-2000 km s$^{-1}$) `mini-BAL' troughs are included (see \citealt{RHH11}
for a full discussion on mini-BAL quasars).\footnote{BAL quasars are,
historically, defined as quasars that exhibit blueshifted absorption due
to the \civ\ doublet at \lala\ 1548.203, 1550.770~\AA\ that is at least
2000 km s$^{-1}$ wide and can extend from $3000$~\kms\
to $25000$~\kms, where 0~\kms\ is at the systemic redshift of
the quasar \citep{WM91} and positive velocities indicate motion toward the observer.
Modifications to this definition have been proposed, e.g. \cite{HA02} and \cite{TH06}.
Herein, we consider absorption at any velocity to be a candidate BAL trough, and we report
widths of confirmed troughs so that others may classify the troughs as they see fit.}
The disk-wind model of luminous Active Galactic Nuclei (AGN)
characterizes BAL features as a result of
material lifted off the accretion disk surrounding the central supermassive
black hole (SMBH) and accelerated by radiation line driving to high outflow
velocities that we observe as blueshifted absorption (e.g. \citealt{MC95},
\citealt{OC10}). Whatever their origin, quasar outflows provide insight
into the physical and
chemical properties
of the central engine of quasars, and may also represent a mechanism
by which SMBHs provide feedback to their host galaxy (e.g.,
\citealt{MA09},
\citealt{AB13},
\citealt{LT14},
\citealt{CA15}).

Variability in the strength (i.e., the depth, width, or outflow velocity profile)
of BALs is a well documented phenomenon
(e.g., \citealt{GB08}, \citealt{HA11}, \citealt{FB13}, \citealt{HB15}).
Specifically, there have been recent studies documenting the
disappearance of BAL
troughs (e.g., \citealt{FB12}) as well as emergence in quasars that were not
classified as having BALs previously (e.g., Rodr\'{i}guez Hidalgo et al.,
in preparation, \citealt{HK08}, \citealt{LH09}).

The cause of BAL-trough variability is still largely debated in the
literature, however, it is likely either due to transverse motion of
absorbing clouds across our line of sight (e.g., \citealt{HA11},
\citealt{MS15}), or due
to changes in the ionization of the absorbing gas (e.g., \citealt{HK08},
\citealt{FB13}, \citealt{WY15}).
Ultimately, it may be a complex mixture of these two scenarios.
Full characterization of BAL variability events (either emergence,
disappearance, or variability in general) would significantly increase
our understanding of both the physics of the quasar's central engine
and the interaction of the quasar with its host galaxy.

In this work we present the discovery of the highest velocity outflow
discovered to date ($\sim$60,000~\kms) at ultraviolet wavelengths,
in the quasar
SDSS~J023011.28$+$005913.6, hereafter J0230 \citep{dr5q}.
The previous highest-velocity absorption identified at ultraviolet wavelengths
in a BAL quasar was at 56,000~\kms\ in PG~2302+029 \citep{jea96}, with the
next highest being at 50,000~\kms\ in PG~0935+417 \citep{RHH11}.\footnote{We
have determined the
features claimed by \cite{fol83} to be O\,{\sc vi} at up to 70,000~\kms\ in
the BAL quasar H~1414+089 are actually S\,{\sc iv}$\lambda$1062 and
S\,{\sc iv}*$\lambda\lambda$1072,1073
absorption in a lower-velocity trough reaching only 28,000~\kms.
That identification is secure because the object's S\,{\sc iv}+S\,{\sc iv}* trough shares
the same distinctive `double-dip' velocity structure as its C\,{\sc iv} and
N\,{\sc v} troughs reaching 28,000~\kms; see Figure 2 of \cite{fol83}.}
Outflows at these extremely high velocities have been previously
observed in X-rays (e.g., \citealt{CB02}, \citealt{PK03},
but see \citealt{ZM15})
and might pose problems for theoretical acceleration models.

We adopt a redshift of $z=2.473\pm 0.001$ for J0230 based on visual inspection
of the Ly$\alpha$, C\,{\sc iii}], and Mg\,{\sc ii} emission lines
and the onset of the Ly$\alpha$ forest.
Our redshift is identical within the errors to the value of
$z=2.4721\pm 0.0005$ given for
this quasar in \cite{bossdr10q}.
We adopt a systematic uncertainty on the redshift of $\pm$0.0044,
or 380 km~s$^{-1}$.
This uncertainty is the difference between the
C\,{\sc iii}] emission-line redshift
and the principal component analysis-based `pipeline'
redshift presented in \cite{bossdr10q}; see that reference for details.
If our adopted redshift is a slight underestimate due to
blueshifting of the emission lines in our spectrum, it is
conservative in the sense that it errs in the direction of
minimizing the observed trough outflow velocities.

J0230 has an apparent magnitude of $g=19.52$ and an absolute
of $M_g=-27$. Because it is undetected in FIRST with an apparent
magnitude of $i=18.76$, it is not radio-loud ($R_i<1$; see
Figure~19 of \citealt{sdss1st}).

In section 2 of this paper, we outline our observations,
data reduction methods, and spectral measurements.
In section 3 we estimate the mass of the black hole.
In section 4 we compare competing models of BAL
variability in the context of our mulit-epoch data.
Finally, we summarize our work in section 5.

Where needed, we adopt a flat cosmology described with
H$_{\circ}=70$ km\,s$^{-1}$\,Mpc$^{-1}$,
$\Omega_{M}$=0.3, and $\Omega_{\lambda}$=0.7.




\section{Spectroscopic Data}

\subsection{Observations}
J0230
was identified as having high-velocity absorption by visual comparison
of its SDSS-I and SDSS-III spectra as part of a search for newly
emerged BAL troughs whose
results will be reported elsewhere. Follow-up
observations were obtained using the Gemini Observatory (see Table~\ref{observ}
for a full list of observations).
The Sloan Digital Sky Survey (SDSS; \citealt{York2000}) obtained two spectra
of J0230
on MJD 52200 and 52942 using a 2.5 meter Ritchey-Chretien telescope
located at the Apache
Point Observatory in New Mexico. We retrieved the fully reduced spectra from
the publicly available Data Release 7 (DR7) quasar catalog \citep{dr7q}.
These spectra
cover the wavelength range  3805$-$9221~\AA\ and 3813$-$9215~\AA, respectively, with
a spectral resolution of $R\sim2000$.
The Baryon Oscillation Spectroscopic Survey (BOSS; \citealt{bossover}), which is part of
SDSS-III, obtained two more spectra of J0230 on MJD 55209 and 55455
using the same telescope as SDSS, but outfitted with a new spectrograph. We retrieved
the fully reduced spectra from the publicly available Data Release 9 (DR9)
quasar catalog
\citep{bossdr9q}. These spectra cover the wavelength range 3574$-$10349~\AA\ and
3594$-$10384~\AA, respectively, with a similar spectral resolution to SDSS.

We observed J0230 on 15 August 2013 (MJD 56519) using the Gemini Multi-Object
Spectrograph (GMOS) on the 8.1 meter Gemini North Cassegrain telescope
located at the
summit of Mauna Kea, Hawai'i. The B600 grating with 600 lines mm$^{-1}$ was used
set at a spectrum central wavelength of 460 nm (the Blaze wavelength is 461nm).
Combined with a 1.0 arcsec wide longslit, the resultant wavelength coverage was
3143$-$6068~\AA\ with a spectral resolution of $R\sim1688$. The total exposure time
was 1200~s. After noting variability of the broad absorption lines in these data,
we observed J0230 on 23 December 2013 (MJD 56649) using GMOS on Gemini North.
The instrument setup was identical as the previous observation, but with a longer
integration time of 2500~s. The resultant wavelength
overage was 3142$-$6077~\AA\ with a spectral resolution of $R\sim1688$. Finally,
after noting further variability in J0230's broad absorption lines, we observed
the object again on 28 January 2014 (MJD 56686) using GMOS on Gemini
South, the twin telescope to Gemini North, located in Cerro Pach\'{o}n, Chile. An
identical instrument setup was used, with a total exposure of 2600~s, resulting
in a wavelength coverage of 3145$-$6077~\AA\ with a spectral resolution of
$R\sim1688$. All three Gemini observations were observed at the parallactic
angle. The data were processed and extracted by standard techniques
using the Gemini IRAF package. The relative fluxes for the three Gemini
spectra were calibrated using
spectrophotometric standard stars; the standard stars were not observed
on the same night as J0230. See Table~\ref{observ} for a list of observations
and the nomenclature we have adopted throughout this paper.

\begin{figure}
\centering
\includegraphics[width=\columnwidth]{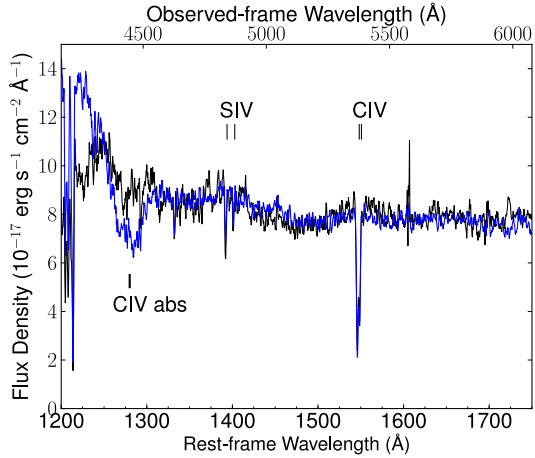}
\caption{Spectra of J0230 at rest-frame wavelengths (bottom scale) and observed
(top scale).
The black spectrum is the mean SDSS spectra
(see \S~\ref{sdssnorm}).
The blue spectrum was taken by BOSS on MJD 55455.
The locations of the \civ\ and \siv\ emission are
labeled (though are weak). In comparing the spectra, a broad and deep trough was
identified at roughly 1262$-$1302~\AA. This trough was identified as highly
blueshifted \civ\ absorption.
This trough is referred to as trough A for the remainder of the paper.
The Flux Density of the BOSS spectrum is artificially scaled up to match
the continuum level of the SDSS spectrum for the purposes of visual comparison.}
\label{discovery}
\end{figure}

In Figure~\ref{discovery} we plot the visual comparison of the mean
SDSS spectrum (black; see \S~\ref{sdssnorm}) and the BOSS2 spectrum
(blue) that led to the identification of an emergent absorbing trough.
The locations of the \civ\ and \siv\ emission are given, though noted to
be very weak (see \S~\ref{description} for further discussion
regarding Weak Line Quasars).
The absorbing trough, which we refer to as trough A for the remainder of the
paper, emerged at some point between the two spectral epochs and spans roughly
1260$-$1300~\AA. We attribute this trough to \civ\ absorption by highly
blueshifted gas outflowing along our line of sight to the quasar at
approximately $\sim$56,000~\kms. We are confident
trough A is not due to blueshifted \siv\ absorption due to the lack of
accompanying \civ\ expected at $\sim$1425~\AA. Further, there is some evidence
that trough A has accompanying \nv\ absorption, which we discuss in
\S~\ref{description}. We note there is a significant change in J0230's spectrum
shortward of trough A, which is attributed to changes in
the \lya+\nv\ complex in that region. Note that those changes do not
affect our measurements on trough A throughout this work.

%
\begin{table*}
\centering
\begin{tabular}{ccccclcc}
\hline \hline
MJD$_{Obs}$ & Rest $\Delta t$ & Rest Day & Plate & Fiber & Origin & SN$_{1675}$ & Name \\
\hline
52200.39 & 000.00 & $-$866.02 &  705 & 407 & SDSS-I & 7.50 & SDSS1                \\ %
52942.34 & 213.63 & $-$652.39 & 1509 & 365 & SDSS-I & 7.57 & SDSS2        \\ 
55208.10 & 652.39 & 000.00 & 3744 & 634 & SDSS-III/BOSS & 12.3 & BOSS1     \\ 
55454.46 & 71.93  &   71.93 & 4238 & 800 & SDSS-III/BOSS & 16.4 & BOSS2        \\ 
56519.53 & 306.67 &  378.06 & ...  & ... & Gemini-North & 22.1 & GEM1 \\ %
56649.21 & 37.33  &  415.39 & ...  & ... & Gemini-North & 23.2 & GEM2   \\
56685.07 & 10.32  &   425.71 & ...  & ... & Gemini-South & 18.0 & GEM3    \\
\hline \hline
\end{tabular}
\caption{Spectroscopic observations of J0230.
Rest $\Delta t$ is the rest-frame time in days elapsed since
the previous observation. Rest Day is cumulative rest days
relative to the first BOSS observation. SN$_{1675}$ is the median
value of the normalized flux divided by the error in the flux
over the spectral range 1650$-$1700~\AA. The final column indicates
how we will refer to each epoch for the duration of the paper.}
\label{observ}
\end{table*}

\subsection{Normalization}\label{norm}
In order to compare the changes we observed in the absorption features,
we normalized each spectrum by a model of the underlying
continuum. First, we smoothed each spectrum using a boxcar average over
5 pixels. Then we identified four regions that appeared unchanged across
all 7 observations; hereafter, these are referred to as normalization
windows. Lastly, we fitted a power-law continuum model to the windows
using a least squares routine.
The normalization windows we used were 1305$-$1330, 1410$-$1420,
1590$-$1620, and 1650$-$1700~\AA.
While there are other normalization windows that could have been
used longward of $\sim1700$~\AA\ for SDSS and BOSS, we refrained
from using them because the Gemini spectra have significantly less
spectral coverage. In order to compare the SDSS/BOSS spectra
to the Gemini spectra we used normalization windows accessible
from all data. We have visually inspected
all our spectra and are confident these windows represent regions
that are unchanged over all epochs of our observations.
We note this is not a normalization in the traditional sense, as
we did not fit the emission features.

Below we describe individual details for normalizing the
SDSS, BOSS, and Gemini spectra.

\subsubsection{SDSS Normalization}\label{sdssnorm}

\begin{figure}
\centering
\includegraphics[width=\columnwidth]{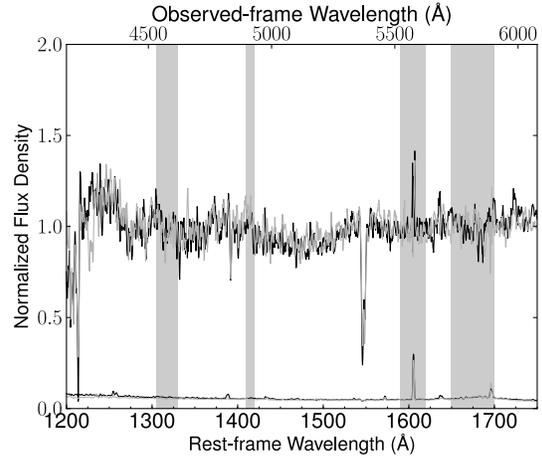}
\caption{The normalized SDSS spectra.
While there are some small differences between the two spectra
(see description in \S~\ref{sdssnorm}), since they do not interfere with
the two troughs we study later we disregard them. As a result, we have
combined the two SDSS spectra together for the analysis throughout the
paper. The normalized error spectra are plotted at the bottom.
The normalization windows are shown as gray regions.}
\label{SDSSnorm}
\end{figure}

Two spectra of J0230 were taken on MJD 52200 and 52942, as part of the
SDSS-I survey.
In Figure~\ref{SDSSnorm}
we present the normalized SDSS spectra; the gray regions indicated the
normalization windows. The normalized error spectra are plotted
along the bottom.
Visual comparison show little difference between the two,
with the small exception of apparent absorption at 1225$-$1235~\AA\
present in the spectrum taken on MJD 52942, but not present in the
previous epoch, taken on MJD 52200. This feature vanished by
BOSS2 and never re-appeared, it is present only in our noisiest spectrum,
and, most importantly, is not related to the two broad troughs that
are the focus of this work. As a result, we do not
consider it in this study.
Other than this feature, there are little differences between the
two SDSS spectra; we
combined them into one (hereafter, 'the SDSS spectrum')
in order to
increase our signal to noise. We adopt an observation date for this
combined spectrum of MJD 52942, that of the latter SDSS observation.
Since no broad absorption is present in either of the SDSS spectra,
we can confidently indicate this date to be the last time we observed no
absorption present.


\subsubsection{BOSS normalization}

\begin{figure}
\centering
\includegraphics[width=\columnwidth]{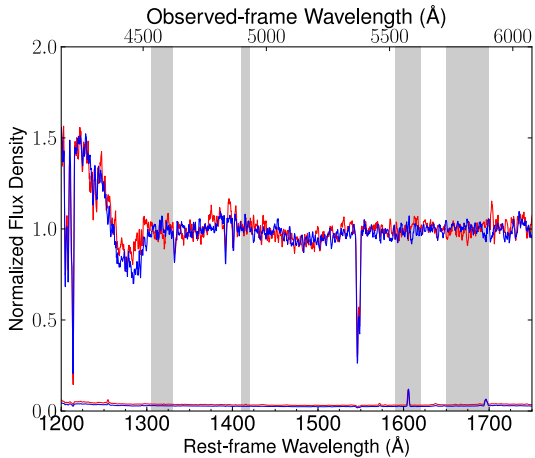}
\caption{The normalized BOSS spectra. Since the high-velocity
absorber at $\sim$1280~\AA\ has changed between the two
epochs, we cannot combine the BOSS spectra. The normalized
error spectrum for both epochs is plotted at the bottom.
The normalization windows are shown as gray regions.}
\label{BOSSnorm}
\end{figure}

The BOSS survey observed J0230 two more times on
MJD 55209 and 55455.
We normalized these two spectra using the same normalization windows
as were used for the SDSS spectra.
In Figure~\ref{BOSSnorm} we plot
the normalized BOSS spectra. In both BOSS epochs, trough A is
present at $\sim$1280~\AA.
The absorption line varies between the two BOSS observations,
thus we did not combine the two spectra as in the case of
the SDSS spectra.


\subsubsection{Gemini Normalization}\label{gemnorm}
\begin{figure}
\centering
\includegraphics[width=\columnwidth]{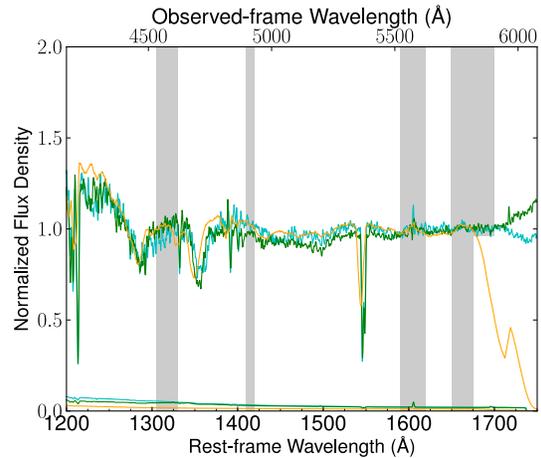}
\caption{The normalized Gemini spectra. The normalization windows
are indicated by the gray regions. The orange spectrum (GEM3) has
slightly less coverage on the red end and thus we changed its
third normalization window to 1640$-$1650~\AA.
The normalized error spectra for all three epochs are plotted
at the bottom.}
\label{GEMnorm}
\end{figure}

Three Gemini spectra were taken on MJDs 56519, 56649, and 56685.
In Figure~\ref{GEMnorm}, all three normalized Gemini
spectra are plotted.
In GEM1 we note the emergence of trough B, a separate medium-velocity
absorber at 1350$-$1360~\AA, which was not present in any of the
SDSS or BOSS spectra.
The Gemini spectrum taken on MJD 56685 (orange)
exhibits less spectral coverage on the red side; the flux falls off
quickly after $\sim$1675~\AA. To account for this, the third
normalization window used for this spectrum was 1640$-$1650~\AA.


\subsubsection{Final Spectra}

\begin{figure}
\centering
\includegraphics[width=\columnwidth]{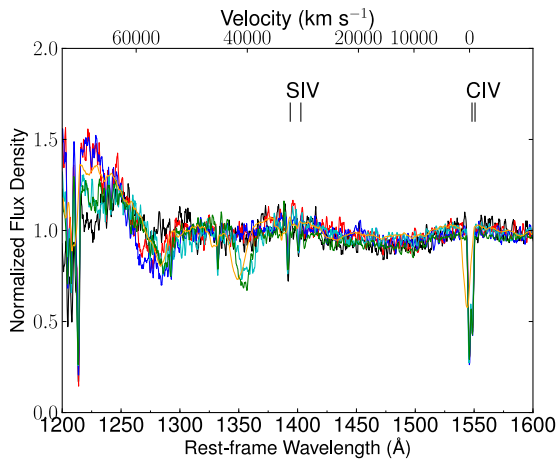}
\caption{All 6 epochs of spectra plotted together. For reference,
the emission features for \siv\ at $\sim1400$~\AA\ and \civ\ at
$\sim1550$~\AA\ are marked, as well as the two troughs `A' and `B' we
observed to emerge during our monitoring campaign. In the legend the
MJD of each observation is indicated as well as the number of rest-frame
days since the previous observation. We also note the presence of
a third mini-BAL feature labeled trough `C' near the systemic
redshift of the quasar. There was no significant change to trough
C through all observations.}
\label{all}
\end{figure}

The final six spectra (note the two SDSS spectra were combined) are
plotted in Figure~\ref{all}. For reference, the emission features for
\siv\ at $\sim1400$~\AA\ and \civ\ at $\sim1550$~\AA\ are marked;
although both emission lines appear to be weak.
In our collected data, we note two
broad absorption features, labeled `A' and `B' in the figure. Trough A
was first observed in the BOSS1 spectrum. At its
widest (BOSS2) trough A spans 40~\AA\ (1262$-$1302~\AA).
Trough B was first observed in GEM1.
At its widest (GEM1) it spans 24~\AA\
(1344$-$1368~\AA). The legend of Figure~\ref{all}
indicates the number of rest-frame days since the previous observation.

\subsection{Summary of Spectral Features}\label{description}

In all spectra we obtained of J0230, the emission features are
relatively weak compared to typical quasars;
specifically, we measured the rest-frame equivalent widths (EWs)
of the emission features:
\lya$+$\nv$=8.0\pm0.1$~\AA, \siv\ $<1.8$~\AA, \civ\ $<2.5$~\AA,
\aiii$+$\ciii\ $=6.1\pm0.2$~\AA, and \mgii\ $=9.8\pm0.9$~\AA.
When there is no apparent emission feature at the
expected location of an ion, we measured the statistical
noise in the spectrum in the ranges provided by the
\citet{VR01} composite quasar spectrum (see Table~2 therein).
Specifically, they measured the \siv\ EW over 1360$-$1446~\AA,
and the \civ\ EW over 1494$-$1620~\AA. For those regions
we measure the statistical noise in our spectra to be
0.60~\AA\ and 0.84~\AA, respectively. The upper limits
quoted above are three times this statistical noise to indicate
the largest possible EW these emission features could have
that would still be statistically below the noise in our data.
Also note that our measurement
of \lya$+$\nv\ is contaminated by the \lya\ forest; the actual
EW is likely larger.
In \citet{LB15}, a Weak Line Quasar (WLQ) is defined as
a quasar whose emission lines have rest-frame
equivalent widths of $<$5~\AA\ (they drew their sample of
WLQs from \citealt{PA10}). While J0230 does not
strictly meet the criterion laid in those works, its
emission features are still far from a typical
quasar's. The original WLQ, PG~1407$+$265,
has emission features with comparable EWs to J0230
(\citealt{MCE95}, see Table~2 therein), as does
the prototypical WLQ, PHL~1811,
which which has the following EWs:
\lya$+$\nv$=15$~\AA, \civ\ $=6.6$~\AA,
\aiii$+$\ciii\ $<$4~\AA, and \mgii\ $=12.9$~\AA\ \citep{LH07a}.
Further, quasars with EWs $<10$~\AA\ investigated so far
have sufficient similarities (e.g., common X-ray weakness)
and can likely be unified as per \citet{LB15}
in a common physical model.
Therefore, we consider J0230 a WLQ.\footnote{WLQs
tend to have blueshifted broad emission lines in the UV,
making systemic redshift determination more challenging than usual.
Our adopted systematic redshift uncertainty of $\pm$380~\kms\
in J0230 is similar to the $+300$~\kms average difference between
redshifts determined by narrow-line studies and those determined
by SDSS for weak line quasars found by \citet{PA10}.}


Also present in all spectra is a narrow \civ\ absorption
feature at $\sim$1550~\AA, very close to
the systemic redshift of
J0230 (also seen in \siv, \cii, \nv, and \lya),
hereafter trough C. There were no dramatic changes in trough C
in our observations.



\begin{figure}
\centering
\includegraphics[width=\columnwidth]{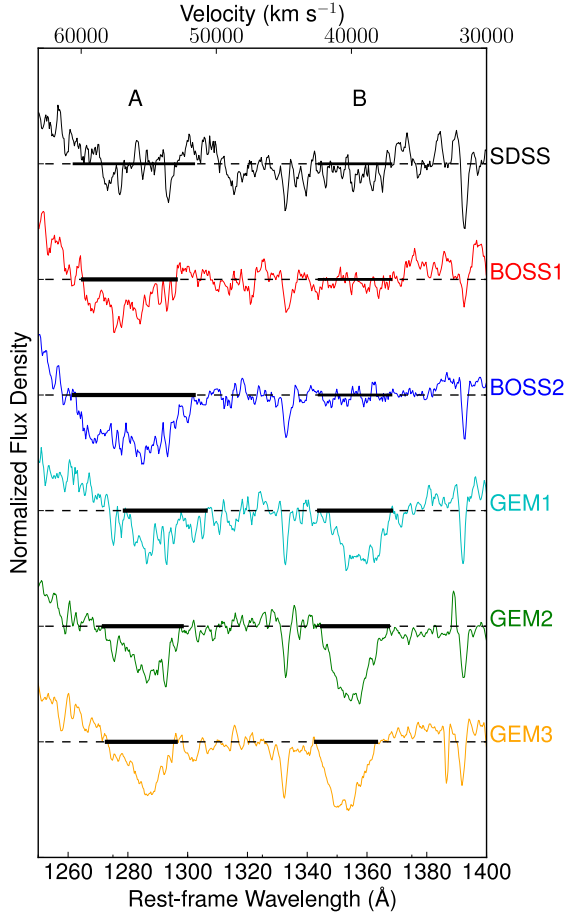}
\caption{Each spectral epoch is plotted centred on the two absorbers,
troughs A and B. We have separated the spectra artificially by 0.5
normalized flux units, with the earliest epoch (SDSS) at the top, and the
most recent (GEM3) at the bottom. The dashed lines indicate the normalized
continuum level for each spectrum. The black bars indicate where we define
the absorption features to begin and end. Note for the SDSS spectrum, there
is no apparent absorption in either troughs A or B. Also note for SDSS,
BOSS1, and BOSS2 there is no apparent absorption for trough B. For these
cases we have placed a slightly thinner black bar across the regions
that represent the widest that trough became. For trough A, this occurs in
BOSS2 and for trough B this occurs in GEM1.}
\label{bothsep}
\end{figure}

The changes in the spectrum are best seen in
Figure~\ref{bothsep}.
In BOSS1 we note the appearance of trough A:
a broad, high-velocity absorber covering the wavelength
range 1260$-$1300~\AA.
Trough A grew to its strongest in BOSS2 by getting both
deeper and wider; these changes were mostly in the
the low-velocity half of the trough, whereas the
high-velocity half of the trough changed less.
In the first Gemini spectrum (GEM1), the high-velocity half of
trough A weakened greatly while its low-velocity half weakened
only somewhat, in comparison to BOSS2.
Between GEM1 and GEM2, trough A strengthened slightly on its high-velocity side.
We note in GEM1 the emergence of trough B,
a second high-velocity absorber in the wavelength
range 1344$-$1368~\AA.
We are confident this absorption is
due to highly blueshifted gas along the line of sight to
J0230. It cannot be due to blueshifted \siv\ absorption
because that would require accompanying \civ\
at $\sim$1500~\AA.
Further,
there is some evidence to suggest there is accompanying
\siv\ and \nv\ at similar outflow velocities (see below)y.
Trough B's low-velocity end remained
relatively unchanged
(though slightly weaker) into GEM2,
while its high-velocity side reached
higher outflow velocities. Finally, between GEM2 and GEM3,
trough A did not change appreciably, while
trough B weakened on its low-velocity side
and its high-velocity edge reached higher outflow velocities.
Trough B also decreased in depth.

\begin{figure*}
\centering
\vspace{-0.4cm}
\makebox[\textwidth]{
\includegraphics[width=1.15\textwidth]{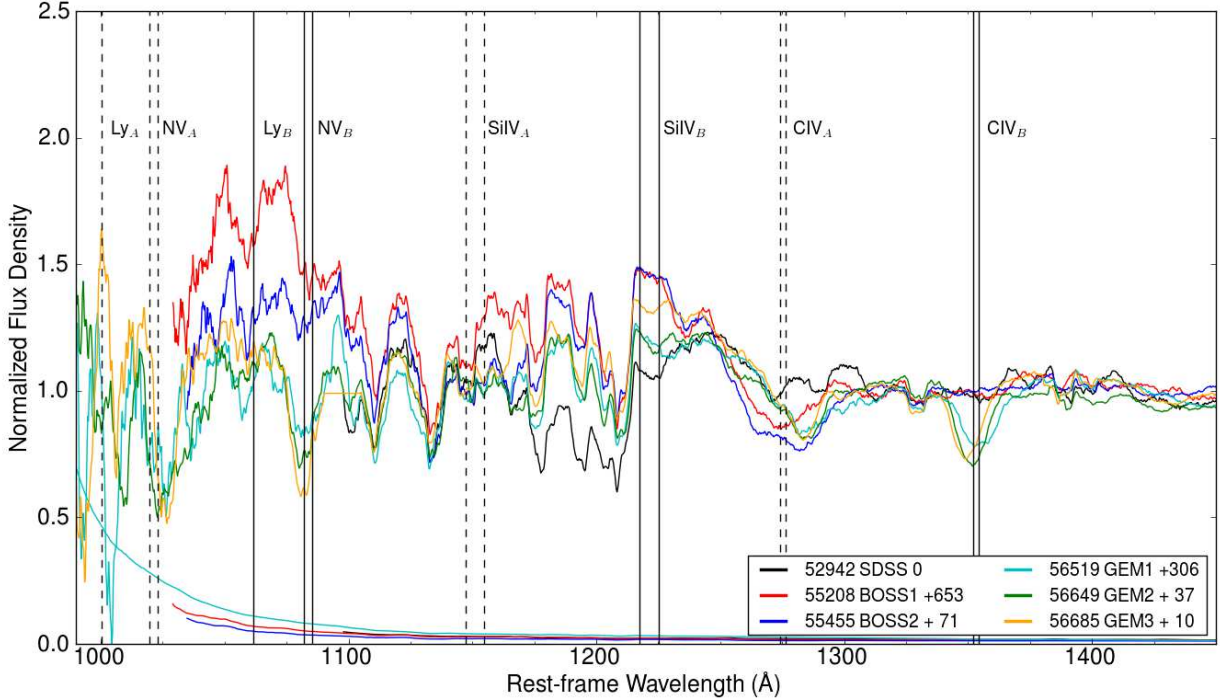}
}
\vspace{-0.6cm}
\caption{Plotted are the 6 normalized spectra with heavier smoothing (boxcar
with window of 25 pixels)
and with a much wider wavelength coverage.
We have marked the location of the trough A \civ\ absorption with a
vertical dashed line; the expected
locations of \siv, \nv, and \lya absorption features that may
accompany trough A's \civ\ are also marked with vertical dashed lines.
The location of trough B's \civ\ absorption is marked with a vertical
solid line, as are the expected locations of this features
possible accompanying \siv, \nv, and \lya.
At the shortest wavelengths of the BOSS1
and BOSS2 spectra
(at $<$1330~\AA\ in the trough A rest-frame),
the spurious broad 'emission' features are due to noise.
It is of note these are the normalized spectra from \S~\ref{norm},
which were created using continuum windows between 1300$-$1700~\AA.
}
\label{wider}
\end{figure*}

The presence of \civ\ absorption can be accompanied by
absorption of one or more other ionic transitions, such as
\siv, \lya, and \nv. We searched for absorption of these ions
that would correspond to the same
outflow velocities as trough A or B. Figure~\ref{wider}
shows all 6 normalized spectra with a much heavier smoothing,
and with a much wider wavelength coverage. We have marked the
observed locations of the \civ\ absorption by trough A
(dashed line) and by trough B (sold line), along with the
expected locations of their accompanying \siv, \lya, and \nv\
absorption. We have plotted the error spectra of the SDSS,
BOSS, and GEM1 spectra along the bottom.
For the purposes of clarity, the spectra were heavily smoothed in
order to see features better in this more noisy part of the spectra.
It is also of note the normalization was not repeated with
new normalization windows in the region from 1000$-$1300~\AA, thus
the relative flux levels are not necessarily accurate. This
is only meant to be a search for possible accompanying absorption.

For trough A, there appears to be no accompanying \siv\ absorption
in any of the spectral epochs we obtained. In searching for
accompanying \lya+\nv, we note that the wavelength coverage
does not extend far enough into the
blue for SDSS, BOSS1, or BOSS2 but does for the three Gemini
spectra. In these latter three epochs there may be \nv, but
no apparent \lya\ is observed.
For trough B, we note the possible presence of
accompanying \siv\
absorption in the three Gemini spectra, however, the absorption
is coincident with the \lya+\nv\ emission systemic to J0230.
Since it is very difficult to disentangle emission from
coincident absorption, we cannot confirm this to be \siv.
The identification is also not certain because a flux deficit was
also seen at that location in the SDSS spectrum, before trough
B appeared. There is probable \nv\ absorption for trough B.


Archival photometry of J0230 is available since it is
located in Stripe 82, a region of sky imaged by SDSS,
multiple times over 7 years \citep{MI10}. We have obtained
the photometry of J0230 from the SDSS archive, however, it is
not concurrent with our spectroscopy. Thus it cannot help us
interpret the spectroscopic variability we observe. J0230 was
too faint for the Catalina Real-time Transient Survey (CRTS).


\subsection{Measurements of Troughs A and B}
We measure the properties of absorption troughs A, B, and C,
such as the equivalent width (EW), the weighted centroid
velocity $v_{cent}$, and the average trough depth, in all
observations in order to compare changes from one epoch
to the next.

In Figure~\ref{bothsep}, the 6 epochs of normalized spectra
are plotted (separated artificially in the y direction).
The bottom x-axis is
the rest-frame wavelength, and the top x-axis is the outflow
velocity relative to \civ~$\sim$1550~\AA. The dashed lines
indicate the continuum for each spectrum. The dark horizontal
lines indicate where we define absorption is present for
troughs A and B (see below on how these were chosen).

To measure the EW from the normalized
spectra we followed equations 1 and 2 in \citet{KB02}, which are,
\begin{equation}
\textrm{EW} = \sum_i \left( 1 - \frac{F_i}{F_c} \right) B_i,
\label{EW}
\end{equation}
and the uncertainty on the EW is,
\begin{equation}
\sigma_{\textrm{EW}} = \sqrt{\left[\frac{\Delta F_c}{F_c} \sum_i\left(\frac{B_iF_i}{F_c}\right) \right]^2 + \sum_i\left(\frac{B_i\Delta F_i}{F_c}\right)^2}.
\label{deltaEW}
\end{equation}
$F_i$ and $\Delta F_i$ are the flux and the error on the flux
in the $i$th bin, $F_c$ is the underlying
continuum flux, $\Delta F_c$ is the uncertainty in the mean of the
continuum flux in the normalization windows, and $B_i$ is the bin
width in units of \AA. In our
normalized spectra, $F_c=1$ and $\Delta F_c$ is calculated using the
windows 1305$-$1330~\AA\ and 1410$-$1420~\AA, which are the two
windows closest to both absorption troughs.
Thus $\sigma_{\textrm{EW}}$ represents the statistical uncertainty
inherent to the spectra. It does not quantify the systemic uncertainty,
which is governed by the placement of the continuum by normalization.

The edges of troughs A and B in a given spectrum were identified by
finding the locations where the flux drops below, and stays below,
the normalized continuum level of $F_c=1$. In Figure~\ref{bothsep},
these edges are represented by black horizontal bars;
in Table~\ref{measure}, $\Delta W$ is calculated using these edges.
We applied Equations \ref{EW} and \ref{deltaEW} to calculate the EW
within the edges found. We note that the placement of the normalized
continuum, and thus the locations of the edges of the troughs, is highly
sensitive to the normalization process.
Further, for trough A, the absorption appears to be truncated by the
\lya+\nv\ emission complex; as a result we consider our EW
measurements to be conservative.

Note for both troughs, some epochs do not exhibit absorption;
both troughs A and B are not present in the SDSS spectrum,
and trough B does not exhibit absorption in the SDSS, BOSS1,
and BOSS2 spectra. For these
cases, we took the largest trough width determined for each trough
and applied it to the unabsorbed spectra. For example, in the case
of trough A, the widest the trough was observed to be was in the
BOSS2 spectrum at $1262-1302=40$~\AA. We applied this range of the
absorption profile in the unabsorbed spectra of SDSS1 and measured
the EW. The resulting value for the SDSS spectrum was $-0.18\pm0.48$,
indicating an EW consistent with zero. More
examples of this can be found in Table~\ref{measure} labeled with
an ellipsis.

We measured the centroid velocity, $v_{cent}$, of the trough
following the definition in \citet{FB13}; it is the mean
of the velocity in a trough where each pixel is weighted by its
distance from the normalized continuum.

The mean depth of the trough was calculated in two ways. First,
we measured $d_{BAL}$ as in \citet{FB13}, which is the mean distance
from the normalized continuum level for each data point in the trough.
Second, we measured $d_{max7}$, which is calculated by sliding a
7 pixel-wide window across the trough and measuring the average depth
over each window. We take the largest value of all these windows as
$d_{max7}$. The uncertainty on the depth calculated as the
uncertainty in the mean of the 7 pixels in the average. We note
that since the our observations were taken with different telescope
and instrument set ups, 7 pixels corresponds to slightly different
resolutions; however, the differences are too small to impact
the measurements. For reference, the 7 pixels covers
approximately 2~\AA, or $\sim$450~\kms\ in all spectra.

\subsubsection{Coordinated Variability}\label{coord}
\begin{figure}
\centering
\includegraphics[width=\columnwidth]{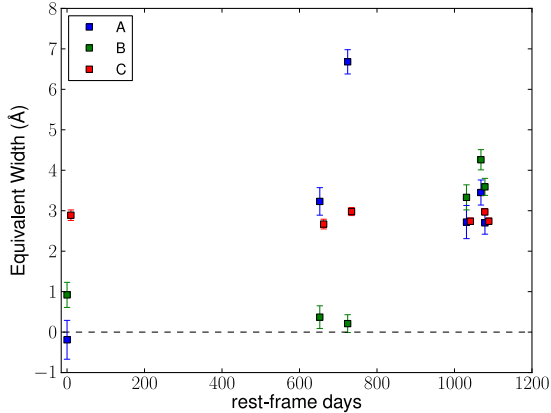}
\caption{The measured equivalent widths (EW) for troughs A (blue),
B (green), and C (red). The trough C points are artificially
shifted to the right by 10 days in order to avoid confusion
with trough A data points.}
\label{EWplot}
\end{figure}

\begin{figure}
\centering
\includegraphics[width=\columnwidth]{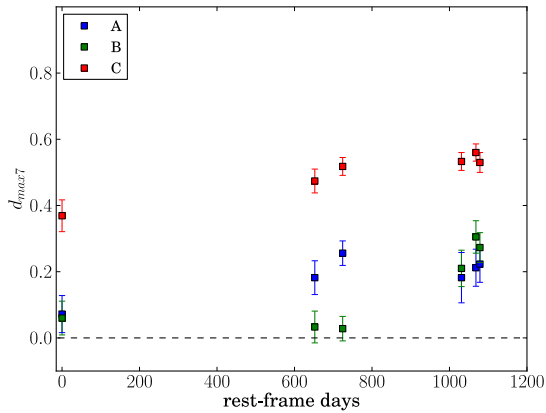}
\caption{The maximum depth
of trough A (blue), B (green), and C (red) as a function
of rest-frame days. $d_{max7}$ represents the lowest
average 7 pixels in a
row for each trough.}
\label{dmax7plot}
\end{figure}

Work on BAL quasar variability indicates troughs from the
same object can vary in coordination with each other,
which can lead to constraints on variability models
(i.e., \citealt{FB12}, \citealt{WY15}; see discussion
in \S~\ref{constraindist} below).
We plot the EW of each trough vs. the rest-frame
time elapsed since the SDSS epoch in Figure~\ref{EWplot}
and $d_{max7}$ for each trough vs. the rest-frame time elapsed
since the SDSS epoch in Figure~\ref{dmax7plot} in order to
investigate how the variability of one trough compares with
the others.
The EW of trough C remains relatively constant over all
epochs. Both trough A and B begin with a very low EW, then
emerge with a sharp and significant increase in later
epochs (BOSS1 for A and GEM1 for B).
There is an interesting pattern in the final three
observations (the Gemini epochs), which occurs
after both troughs have emerged and are established:
the EW for both trough A and B increases for GEM2
and then returns to the same value it was in GEM1
for GEM3. This pattern could be interpreted as absorption
from two physically distinct clouds varying in a
coordinated fashion (for reference, the time frame from
GEM1 to GEM3 is 47 days).
However, the uncertainties on
our EWs are of similar scale 
to the amount of variability we are referring to in the
Gemini epochs. Thus, this pattern does not represent
statistically significant coordination in variability.

\subsubsection{BALnicity Index}\label{BI}

For comparison to other BAL-quasar studies, we have measured the BALnicity
Index (BI) of J0230. We calculate the Absorption Index (AI$_{450}$),
defined in \cite{HA02},
following:
\begin{equation}
{\rm AI}_{450} = \int_0^{v_{high}}\left( 1 - \frac{f(v)}{0.9} \right) C' dv.
\end{equation}
where $f(v)$ is the normalized flux density as a function of velocity, and $C'$
is equal to 1.0 within a trough if the trough is wider
wider
widerthan 450~\kms,
otherwise it is set to 0.0.
The integration begins at $v=0~\textrm{\kms}$ relative to the systemic
velocity of the quasar and runs through the highest velocity at which
absorption is present.

We also measure the modified BALnicity index, BI$^*$, following:
\begin{equation}
{\rm BI}^* = \int_{v_{low}}^{v_{high}}\left ( 1 - \frac{f(v)}{0.9} \right )C dv.
\end{equation}
which is a modification of the original BI defined in \cite{WM91} that imposes no constraint on the outflow velocity.
BI$^{*}$ has no formal limit on the minimum and maximum absorbing
velocities.
The quantity $C$ is equal to 1 only when the quantity in parentheses is
greater than zero for more than 2000~\kms,
otherwise it is set to 0.0.

In Table~\ref{BALnicity}, we list the BALnicity indexes calculated using both methods.
The total index value is measured over $v_{low}>0$ and $v_{high}<60,500$ km s$^{-1}$,
however, we also provide the individual contributions of each trough in the spectra.
Note that for AI$_{450}$, trough C contributes to the total index, but for BI$^{*}$
it does not.

\begin{table*}
\centering
\begin{tabular}{cccccccc}
\hline \hline
 & Rest $\Delta t$ & EW$\pm\sigma_{EW}$ & $\Delta$w & $\Delta$v & $d_{max7}$ & $v_{cent}$ & $d_{BAL}$ \\ 
\hline
Trough A & (days) & (\AA) & (\AA) & (km/s) & & (km/s) & \\
\hline
SDSS & 000.00 &$-$0.19$\pm$0.48 & \ldots & \ldots & 0.07$\pm$0.06 & \ldots & \ldots \\
BOSS1& 652.39 &   3.23$\pm$0.34 & 31 & 7002 & 0.18$\pm$0.05 & 56496 & 0.10 \\
BOSS2& 71.93  &   6.68$\pm$0.30 & 40 & 9028 & 0.26$\pm$0.04 & 56004 & 0.17 \\
GEM1 & 306.67 &   2.72$\pm$0.41 & 27 & 6063 & 0.18$\pm$0.08 & 53769 & 0.10 \\
GEM2 & 37.33  &   3.45$\pm$0.31 & 26 & 5860 & 0.21$\pm$0.06 & 55020 & 0.13 \\
GEM3 & 10.32  &   2.70$\pm$0.28 & 23 & 5185 & 0.22$\pm$0.06 & 55101 & 0.12 \\
\hline
Trough B & & & & & & & \\ 
\hline
SDSS & 000.00 & 0.92$\pm$0.31 & \ldots & \ldots & 0.06$\pm$0.05 & \ldots & \ldots \\
BOSS1& 652.39 & 0.37$\pm$0.28 & \ldots & \ldots & 0.03$\pm$0.05 & \ldots & \ldots \\
BOSS2& 71.93  & 0.21$\pm$0.22 & \ldots & \ldots & 0.03$\pm$0.04 & \ldots & \ldots \\
GEM1 & 306.67 & 3.33$\pm$0.31 & 24 & 5214 & 0.21$\pm$0.05 & 39212 & 0.14 \\
GEM2 & 37.33  & 4.26$\pm$0.25 & 22 & 4780 & 0.31$\pm$0.05 & 39726 & 0.11 \\
GEM3 & 10.32  & 3.59$\pm$0.21 & 20 & 4352 & 0.27$\pm$0.04 & 40224 & 0.10 \\
\hline
Trough C & & & & & & & \\
\hline
SDSS & 000.00 & 2.89$\pm$0.13 & 8 & 1549 & 0.49$\pm$0.04 & 87 & 0.37$\pm$0.05 \\
BOSS1& 652.39 & 2.67$\pm$0.12 & 8 & 1550 & 0.47$\pm$0.04 & 78 & 0.33$\pm$0.05 \\
BOSS2& 71.93  & 2.98$\pm$0.10 & 10 & 1935 & 0.52$\pm$0.03 & 68 & 0.30$\pm$0.05 \\
GEM1 & 306.67 & 2.74$\pm$0.08 & 7 & 1356 & 0.53$\pm$0.03 & 163 & 0.39$\pm$0.04 \\
GEM2 & 37.33  & 2.97$\pm$0.07 & 7 & 1356 & 0.56$\pm$0.02 & 125 & 0.42$\pm$0.04 \\
GEM3 & 10.32  & 2.74$\pm$0.08 & 7 & 1356 & 0.53$\pm$0.03 & 212 & 0.39$\pm$0.04 \\
\hline \hline
\end{tabular}
\caption{Measurements made on trough A, B, and C.
The '\ldots' indicate where no absorption is visible in the spectrum.
Values of EW for these cases used the widest possible $\Delta W$ the
trough was observed to reach (BOSS2 for trough A, GEM1 for trough B).}
\label{measure}
\end{table*}


\begin{table*}
\centering
\begin{tabular}{ccccccccc}
\hline \hline
 & AI$_{A}$ & AI$_{B}$ & AI$_{C}$ & total AI & BI$^*_A$ & BI$^*_B$ & BI$^*_C$ & total BI$^*$ \\
\hline
SDSS   & 0.0 & 0.0 & 477$\pm$3 & 477$\pm$3 & 0.0 & 0.0 & 0.0 & 0.0 \\
BOSS1  & 152$\pm$5 & 0.0 & 437$\pm$2 & 589$\pm$5 & 0.0 & 0.0 & 0.0 & 0.0 \\
BOSS2  & 746$\pm$6 & 0.0 & 490$\pm$2 & 1236$\pm$6 & 561$\pm$5 & 0.0 & 0.0 & 561$\pm$5 \\
GEM1 & 103$\pm$6 & 323$\pm$6 & 455$\pm$2 & 880$\pm$8  & 0.0 & 115$\pm$4 & 0.0 & 115$\pm$4 \\
GEM2 & 293$\pm$6 & 547$\pm$6 & 491$\pm$2 & 1331$\pm$9 & 74$\pm$4 & 242$\pm$4 & 0.0 & 316$\pm$5\\
GEM3 & 210$\pm$6 & 433$\pm$5 & 453$\pm$2 & 1096$\pm$7 & 35$\pm$3 & 155$\pm$3 & 0.0 & 190$\pm$4\\
\hline\hline
\end{tabular}
\caption{The BALnicity was calculated using two different definitions:
AI$_{450}$ and BI$^{*}$ (see \S~\ref{BI}). We calculated the total index
over a velocity range of $v_{low}>0$ and $v_{high}<60,500$ km s$^{-1}$.
We also calculated the individual contributions to the index by each
trough in the spectra. In the case of AI$_{450}$, trough C contributed
to the measurement;
for completeness, we provide its index measurement.
In BI$^{*}$, trough C did not contribute to the total. Note the
uncertainties quoted here are statistical only. Systematic uncertainty
introduced by the placement of the continuum is not taken into account.
A reasonable continuum uncertainty of $\pm5$\%\ translates to a
BALnicity Index uncertainty of $\pm5$\%/$d_{BAL}$.}
\label{BALnicity}
\end{table*}

\section{Black Hole Mass Estimate}\label{massest}

To estimate the mass of the SMBH in J0230, we used the
velocity dispersion of the \mgii\ \lam~2796, 2803~\AA\ emission line.
A full description of this technique can be found in \citet{RafieeH11}.
Equation 9 of that work is
\begin{equation}
M_{BH}/M_\odot =30.5[\lambda L_{3000}/(10^{44} {\rm ~erg~s^{-1}})]^{0.5} \sigma^2
\end{equation}
where $L_{3000}$ is the observed monochromatic luminosity at
3000~\AA\ rest-frame, $\lambda=3000$~\AA, and $\sigma$ is the intrinsic line
dispersion of the \mgii\ emission line in \kms.
There is intrinsic scatter of $\pm 0.15$~dex ($\pm 35$\%)
and systematic uncertainty of $\pm 0.10$~dex ($\pm 24$\%) in this equation.

\begin{figure}
\centering
\includegraphics[width=\columnwidth]{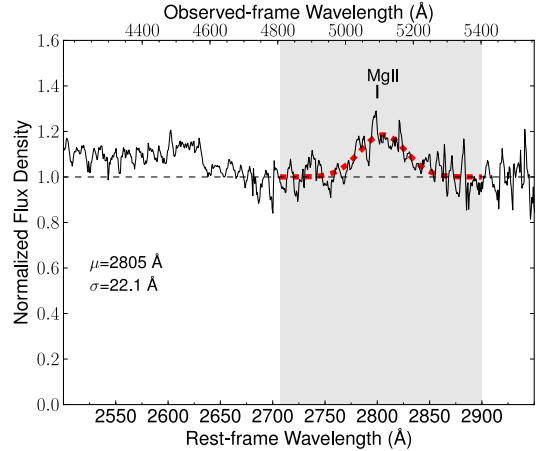}
\caption{Combined BOSS spectra (black), continuum fit (dashed black), and
Gaussian fit (dashed red) to the \mgii\ emission feature at
$\sim2800$~\AA. The fit was applied only to the data in
grayed out region. The best-fit Gaussian parameters to the data are shown
in the lower left.}
\label{mgiifit}
\end{figure}

The two BOSS spectra of J0230 represent the best coverage we
have of that wavelength regime. We combined the two BOSS spectra with a
weighted mean, and then
fit a line to the continuum using windows 2650$-$2700~\AA, 2900$-$3000~\AA.
After fitting and removing the continuum, we fit a Gaussian to the
remaining \mgii\ emission in the region 2700$-$2900~\AA. In Figure~\ref{mgiifit}
the fitted Gaussian is plotted over the normalized BOSS spectra. The
best-fit
parameters were $\mu=2805$~\AA,
and $\sigma=22.1$~\AA. The standard deviation in the Gaussian fit
indicates the velocity dispersion of the \mgii\ emission feature, which
is caused by the Doppler broadening of an AGN broad line region orbiting
the SMBH. We convert $\sigma =22.1$~\AA\ $=$ 2370~\kms.

To calculate the quasar luminosity we used
\begin{equation}
\lambda L_{3000}=4{\rm \pi} D_L^2 f_{3000}\times 3000(1+z),
\end{equation}
where $D_L$ is the luminosity distance, $f_{3000}$ is the observed
flux density at rest-frame 3000~\AA, and z is the redshift.
We measured $f_{3000}$ from the combined BOSS spectrum
to be $f_{3000}=3.35\times10^{-17}$ erg s$^{-1}$ cm$^{-2}$ (observed \AA)$^{-1}$.
For our adopted cosmology, the luminosity
distance to $z=2.473$ is $D_L=2.01 \times 10^{10}$~pc,
or $6.21 \times 10^{28}$~cm.
Therefore, we have $\lambda L_{3000}=1.69\times 10^{46}$ erg s$^{-1}$.
Finally, we calculate the mass of the SMBH to be
$M_{BH}/M_\odot = 2.2 \times 10^{9}$ with an intrinsic scatter
of $\pm 0.8 \times 10^{9}$ ($\pm$35\%).

For a SMBH of this mass the Eddington Luminosity is
$L_{Edd}=3.45\times 10^{47}$ erg~s$^{-1}$. Using a bolometric
correction of $BC_{3000}=5$ derived in \citet{gtr06},
this quasar has an estimated $L_{Bol}=8.45\times 10^{46}$ erg s$^{-1}$,
and therefore this quasar has an estimated
$f_{Edd}\equiv L_{Bol}/L_{Edd}=0.25$.
Such a black hole has $R_{Sch}=6.6\times 10^{9}$~km.

It is worth pointing out that \citet{PA10} present some
evidence to suggest that some \mgii\ emission lines of WLQs
could exhibit non-virialized behaviour (namely, the emission
feature is blueshifted from the systemic redshift, though only
by 360~\kms\ on average; see \S\ 5.2 and 6.1 of that work).
We see no such evidence of a non-virialized \mgii\ emission
feature in J0230: a single Gaussian function fits the
emission line well, its peak is actually redshifted
by $\sim510\pm 380$~\kms\ from the position of \mgii\
expected from the composite spectrum of \citet{VR01}.

Moreover, we have calculated black hole masses using
the dispersion in the \mgii\ emission line for the objects
from \citet{PA10}. We find that the resulting masses are
larger than the black hole masses they
calculate using the dispersion in \hb\ by only a factor of
two. A deviation of that factor is not statistically significant
given the uncertainties on our black hole mass estimate.

\section{Discussion}
As mentioned in the introduction, 
broad absorption trough variability in quasars
can be explained by
transverse motion of the absorbing clouds across the line of sight
to the accretion disk, or by changes in the ionization parameter
of the absorbing cloud, or by a combination of these.
Here we analyze two possibilities
individually laying out constraints where possible.
Note that in the end, the range of possible locations for the gas is
large enough to preclude useful constraints on the kinetic luminosity
of the outflow \citep{DB10}, especially since the solid angle
covered by this extremely high velocity outflow is unknown.

\subsection{Pure Transverse Motion Variability Model}\label{transmotion}
In the transverse-motion model it is assumed the absorption parameters
of the cloud of gas are unchanged, and all changes to the EW, the
velocity profile, and the maximum depth of the trough can be
explained by an absorbing cloud moving to cover more or less of the
accretion disk. Any evolution of an absorption feature (i.e., an
emergent trough) can be explained as long as the constraints from
timescales yield plausible transverse velocities.

The transverse velocity of an absorbing cloud across the line
of sight is derived by dividing the distance the cloud travels
by the travel time it took to get there, the latter of which is
time between successive observations.
In order to measure the distance covered by an absorbing cloud
between those observations we must both estimate the size of
the continuum region it is traversing, and also model the
relative sizes and shapes of the cloud and continuum region.

We approximate the continuum region to be represented by the
$\alpha$-disk model presented in \citet{SS73}, hereafter SS73,
with the following model parameters. We set $\alpha=0.1$,
a free parameter in the model that governs the amount of
accretion as a result of turbulence, typically $0<\alpha<1$.
We assume a non-spinning black hole, which leads to an
accretion efficiency of $\eta=0.0572$. Given these parameters,
the rate of mass accretion onto the black hole would be
$\dot{m}\equiv f_{Edd}/\eta=0.25$.
Using an accretion disk defined by these parameters, we
can estimate
$D_{95}(1320)$, the continuum diameter within which 95\%
of the 1320~\AA\ continuum is emitted. We use the 1320~\AA\
continuum, which is the region in between troughs A and B,
because it allows us to use a conintuum region that is the
same size for both troughs; we note the size of the accretion
disk would only change a small percentage if using the
trough A or B centroid wavelengths.
We find $D_{95}(1320)=63~R_{Sch}$, therefore,
$D_{95}(1320)=4.2\times 10^{11}$~km.
That gives a light-crossing time of $1.4\times 10^6$~s = 16 days.

However, accretion-disk sizes inferred from gravitational-microlensing
studies and photometric-reverberation studies
(e.g., \citealt{MK10}, \citealt{BP11}, \citealt{JM12}, \citealt{EG15})
are approximately a factor of four larger than the theoretical
size predicted in the SS73 $\alpha$-disk model
(see a full discussion in \citealt{HN14}).
Therefore, we increase our estimated continuum-source diameter
by a factor of four, to
$D_{95}(1320)=252~R_{Sch}=1.7~\times 10^{12}$~km.
The uncertainty in this number is likely a factor of two.
A disk that size has a light-crossing time of
$5\times 10^6$~s = 64 days.

With the estimated size of the emitting
region, and, given some simple models of clouds
moving into or out of the line of sight of an
emitting region, we can estimate a maximum and minimum transverse
velocity of an absorption cloud that would be responsible
for the emergence and variability of troughs A and B.

The most dramatic change we observed in the absorption depth
of J0230 occurred in trough B when it emerged between the BOSS2
and GEM1 observations; the change in depth was
$\Delta d_{max7}=0.21-0.03=0.18$ over a period of 307 rest-frame
days.
As per the transverse-motion model, if we consider this change
in depth to be entirely due to more of an optically thick absorbing cloud
moving into the line of sight to the emitting region,
it suggests that
over 307 rest-frame days, the emitting region went from having
3\%\ of its flux blocked to having 21\%\
blocked, or a 21\%\ covering fraction, $C$.
Note in order for changes
in absorption depth to equate to changes in covering fraction
we are assuming the lines are optically thick.
(If the lines are optically thin, the absorber must reach
a larger covering fraction of the emission region in the
same time span, requiring even higher transverse velocities.)

In \citet{CH13}, two simple models were proposed for clouds
crossing the emitting region (see Figure~14 therein).
The first scenario is the `crossing disks' model, where the
absorbing cloud is projected on the sky as a circle (or a
disk) and is crossing a circular emitting region (where the
emitting region appears much larger than the absorbing cloud).
In the second scenario the absorber is much larger than the
background emitter it is traversing; this is the `knife-edge'
model. As mentioned above, the crossing speeds in these two
scenarios are measured by dividing the distance traveled by
the time it took to travel there. The change in covering
fraction, $\Delta C$, is the fraction of the
emitting region the absorber crosses in the time-frame
between observations. Therefore, in the `crossing disks'
scenario, the minimum distance traveled by the gas responsible
for the emergence of trough B is $\sqrt{\Delta C}D_{95}(1320)$,
and the crossing time is $\Delta t=307$ days. Therefore,
$\sqrt{0.18}\times 64$ light-days $=27$ light-days in 307
days. Therefore the transverse speed is 26,400~\kms. However,
if we assume the cloud has traversed to the exact opposite
side of the emitting region, the distance traveled is the
complete 64 light-days in 307 days. This results in a
transverse velocity of 62,500~\kms. In the `knife-edge'
scenario, the distance traveled is
$\Delta C~D_{95}(1320)=12$
light-days in 307 days. This equates to 11,700~\kms. Thus,
given the above two scenarios, we can place the transverse
velocity of a cloud responsible for the emergence of trough B
in the range $11,700 < v (\textrm{\kms}) < 62,500$.
For trough A,
the most dramatic change in absorption depth also occurred
during its emergence, which was between SDSS and BOSS1;
the change in depth was
$\Delta d_{max7}=0.18-0.07=0.11$ over a period of 652
rest-frame days. Applying the same relations as above we
can place the transverse velocity of a cloud responsible for the
emergence of trough A in the range
$3,200 < v (\textrm{\kms}) < 29,500$.

While these two models can be useful in interpreting observations
in a campaign with two epochs, our unique dataset consists of six
epochs. Analyzing the behaviour of the absorption features over all
six epochs allows us to test the predictive power of the above two
scenarios. For instance, trough B was consistent with zero absorption
in the SDSS, BOSS1, and BOSS2 observations (see Table~\ref{measure}).
The trough appeared between the observation of BOSS2 and GEM1,
which was over a time period of 307 days, then for the next 2
observations (GEM2, and GEM3) the trough remained close to the same
depth and EW (within the uncertainties). Assuming an absorber is
moving at a constant velocity transverse to the line of sight, the
above behaviour rules out the `knife-edge' scenario, which would
only cover more area as time goes on.

\begin{figure}
\centering
\includegraphics[width=\columnwidth]{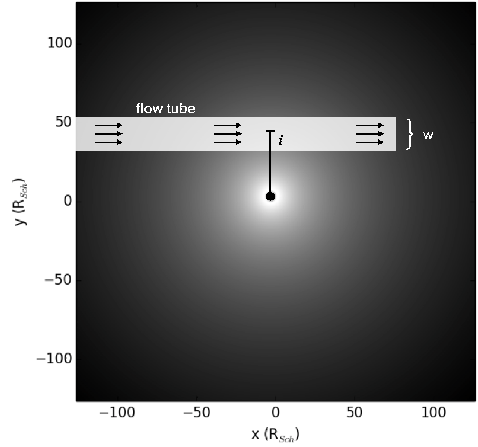}
\caption{An example of a flow-tube traversing a simulated emitting
region of an accretion-disk. The logarithm of the luminosity of the disk
is represented by the gray scale. Over-plotted is an example of a
flow-tube traversing
the disk, which would serve to cover some of the light, creating
absorption. In this representation, the tube extends infinitely to
the left but terminates at the right edge drawn. The width of the tube
is w and the impact parameter relative to the centre of the accretion
disk is i. Note there is a region at the very centre occupied by the
black hole where no luminosity is observed.}
\label{flowtube}
\end{figure}

If we assume the emitting region has a uniform flux across its
area (as \citealt{CH13} does), then the `crossing disk' scenario
can explain trough B's behaviour. However, research into the
theoretical understanding of accretion disks - through the work of
SS73 and \cite{DA11} (among others) - indicates the emitting region
is unlikely to be homogeneous. If we assume the emitting region is
more like a SS73 disk, where the majority of the flux is concentrated
toward the centre of the emitting region, we can also rule out the
`crossing disks' scenario.\footnote{see Fig.~\ref{flowtube}
for an example of the luminosity gradient of a SS73 accretion-disk;
this figure will be discussed in more detail later}
A crossing disk of fixed size traversing a SS73 accretion disk
at a constant velocity would produce an increasing amount of
observed covering fraction as it moved across the first half of
the disk, but then a decreasing amount of covering as it traversed
the second half of the disk. If trough B appeared in GEM1 as a
result of transverse motion, we would have expected to see the depth
of the absorber decrease appreciably in the subsequent observations
of GEM2 and GEM3. Since this is not the case, the `crossing disks'
scenario is unlikely to be the correct interpretation of the
variability of trough B.

Over 6 epochs, the nature of trough A's variability also rules out
the `knife-edge,' but agrees with the augmented
`crossing disks'+SS73 scenario. Specifically, there was no measured
absorption in SDSS after which there was an increase in absorption
in BOSS1 which continued to increase in both depth and EW into BOSS2.
Then by GEM1 through GEM3, both the depth and EW returned back to
values similar to those measured in BOSS1. This is consistent with
a cloud smaller in angular size than the emitting source traversing
into the line of sight for BOSS1, crossing the central portion of
the disk leading to a the measurements of BOSS2, continuing on to
the second half of the disk for GEM1 and GEM2, but has not reached
the other side yet as there is still measured absorption in GEM3. If
we apply the relations from the `crossing disks' scenario above, the
velocity range this absorber would have is
$10,000 < v~(\textrm{\kms}) < 18,000$.
At 18,000~\kms\ we expect trough A to disappear completely
approximately 350 days after the GEM3 observation.
At 10,000~\kms\ we expect it to disappear
approximately 1,500 days after the GEM3 observation.

There is one plausible scenario of transverse motion that can
match the observed depth changes in trough B in the context
of a SS73 disk: a `flow-tube' (similar to that proposed by
\citealt{AK99}; see their Figure~10). In Figure~\ref{flowtube},
we have plotted a log-luminosity map of an accretion-disk emitting
at 1320~\AA\ powered by a SMBH equal to that of J0230 (see
\S~\ref{massest}). The emitted light is much more concentrated
towards the centre (though note there is a region occupied by the
black hole where no emission is observed). We have plotted over
top of the map an example of our proposed flow-tube scenario.
The tube is traversing the continuum region at some impact parameter,
i, away from the centre, and has some width, w. The tube extends
infinitely to the left in this figure. We note that our flow-tube
geometry and dynamics differ from that proposed in \cite{AK99}.
Specifically, we have chosen a flow-tube that is homogeneous from
centre to edge and is in the midst of establishing itself along
our sight-line before settling into a long-term configuration as
discussed in \citet{AK99}.

If a flow-tube similar to the one shown in Figure~\ref{flowtube}
were to move across the emitting region of J0230, it would serve
to create a sharp increase in absorption as it crossed close to the
centre of the disk, but due to there being very little flux at the
edges of a SS73 disk, not much more coverage would occur as it
traversed the second half of the continuum region. This geometry
would match the behaviour we see in the variability of trough B.

We have investigated whether a flow-tube of this nature could
successfully reproduce the variability in trough B, and at what
velocities it could do this, by simulating flow-tubes of various
widths and impact parameters traversing a SS73 disk, measuring how
much flux is covered as a function of distance across the disk the
simulated flow-tubes produce, and then attempting to match the
observed covering fractions for trough B to the simulated covering
fraction vs. distance generated by the flow-tubes. Referring to
Figure~\ref{flowtube}, traversed distance is measured along the
x-axis of the disk, and the direction of motion of each simulated
flow-tube is from the negative x direction towards the positive
x direction.


\begin{figure}
\centering
\includegraphics[width=\columnwidth]{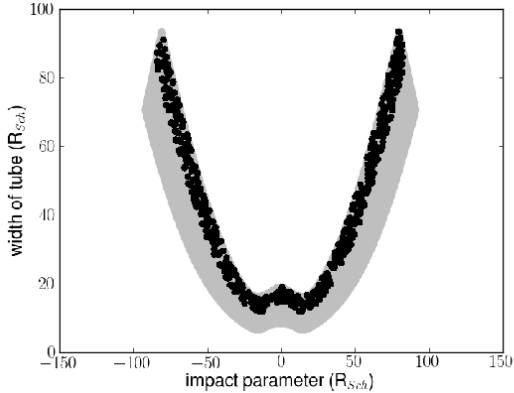}
\caption{Width of flow-tube vs. the distance from centre of
accretion disk the flow-tube traverses (impact parameter). It is
plotted in units of $R_{Sch}=6.6\times 10^{9}$~km. The gray
region represents all possible combinations of flow-tubes that
resulted in a final covering fraction between 15\%\ and 30\%.
The black points are the combinations of parameters that not only
matched all covering fractions in our observations, but did
so within the observation time constraints. The x-axis is
plotted is distance from centre of tube to centre of
accretion disk, where positive and negative values represent
opposite sides of centre.}
\label{vimp}
\end{figure}

\begin{figure}
\centering
\includegraphics[width=\columnwidth]{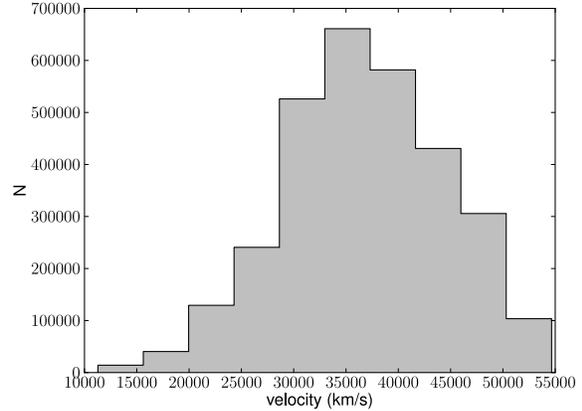}
\caption{The range of possible velocities of a flow-tube traversing
the emitting region of J0230. These were determined by simulating
flow tubes of various widths and impact parameters across a SS73
disk scaled to match J0230's mass and monochromatic luminosity
at 1320~\AA. In
order to be a plausible velocity, the tube must recreate the covering
fraction at each spectral epoch, given one velocity, as well as
match the time between observations.}
\label{vhist}
\end{figure}

Matching the observations to our simulations was done via the
following prescription:
A given simulated flow-tube has covering fraction as a function
of x, $C(x)$. We search for a distance across the accretion disk,
$x_0$, that matches the covering fraction for BOSS2, namely
$0< C(x_0) < C$(BOSS2)+1$\sigma$, which is the last time trough B
was measured to have a depth consistent with zero. When found, we
go searching for the next closest $x_1$ that satisfies
$C(GEM1)-1\sigma < C(x1) < C(GEM1)+1\sigma$. We calculated the
velocity, $v$, required to cover the distance from $x_0$ to $x_1$,
given the known time between successive observation (307 days from
BOSS2 to GEM1). We then searched for the next $x_2$ that satisfies
$C(GEM2)-1\sigma < C(x_2) < C(GEM2)+1\sigma$. When a match is found,
we use the simulated distance from $x_1$ (GEM1) to $x_2$ (GEM2),
and the velocity the flow-tube is moving at, $v$ calculated above,
to determine the length of time it would take for the flow-tube to
cover the distance $x_1-x_2$. If the time is equal to the time
between GEM1 and GEM2 observations (37 days) then we continue the
search to see if GEM3 also matches. We look for $x_3$ that satisfies
$C(GEM3)-1\sigma < C(x_3) < C(GEM3)+1\sigma$. Similar to above,
we use the distance from $x_2$ (GEM2) to $x_3$ (GEM3), and the $v$
above to determine the length of time it would take for the
flow-tube to cover that distance. If that time is equal to the time
between GEM2 and GEM3 observations (10 days) then we have found a
combination of width and impact parameter for a simulated flow-tube
that matches the variability in covering fraction as well as the
time between successive observations.
In Figure~\ref{vimp}, we have plotted the parameter space of
width vs. impact parameter that we investigated with the simulated
flow-tubes. The gray region displays the combinations of parameters
that resulted in a flow-tube's final covering fraction (after it
had completely traversed the disk) between 15\%\ and 30\%, which is
a healthy margin for the GEM3 covering fraction. The black points
represent the combinations that fit the variability of BOSS2 through
to GEM3. In Figure~\ref{vhist} we plot a histogram of all possible
velocities we determine from the above analysis. The mean velocity
of the distribution is $36,800$~\kms\ with a range spanning
$8,000 < v~(\textrm{\kms}) < 56,000$.

Trough A could still be explained as a flow tube, but a simple
flow tube model is not consistent with its $d_{max7}$
and EW variations.
The best fit despite those variations would yield a slower
transverse velocity because the time over which the biggest change
occurred (SDSS-BOSS1) is larger than for trough B. Note that a
slower transverse velocity is consistent with trough A's higher
line-of-sight velocity, as gas which is closer to terminal
velocity is likely farther from the quasar with lower
transverse velocity due to angular momentum conservation.

In summary, we have found that pure transverse motion can
plausibly explain the variability of both trough A and B
over all 6 epochs of observation in our dataset. Trough A
is best explained by a `crossing disks' traversing a SS73 disk
at velocities between $10,000 < v~(\textrm{\kms}) < 18,000$.
This model and velocity range allow us to predict trough A
will disappear between $350 < t~(\textrm{days}) < 1,500$ after
our last observation (GEM3).
Trough B is best explained by a flow-tube that has recently
moved into the line of sight, travelling in the velocity range
$8,000 < v~(\textrm{\kms}) < 56,000$. In this scenario, we have
no constraint on how far a flow-tube extends, and thus cannot
predict when trough B will disappear.

\subsubsection{Constraining Distances}\label{constraindist}

If we assume the absorbers responsible
for both trough A and B have reached maximum velocity,
have transverse velocities $v_{trans}$ and
we are observing them at some current distance $r_{C}$ from the BH,
but was launched from
a circular orbit at a distance $r_{L}$, which has an orbital velocity
$(GM_{BH}/r_{L})^{0.5}$, then we can constrain
both $r_{C}$ and $r_{L}$ using our observed velocities.
From conservation of angular momentum for a gas parcel of mass $m$, we have
\begin{equation}
m\times r_{L}\sqrt{\frac{GM_{BH}}{r_{L}}}=m\times r_{C} v_{trans}.
\end{equation}
Thus the BAL gas transverse velocity is $v_{trans} = \sqrt{GM r_{L}/r^2_{C}}$
(ignoring any transverse component of its velocity away from the black hole
across our line of sight to the continuum source).
The final radial velocity is $v_\infty = F\sqrt{GM / r_{L} } $
where the scaling factor $F$ is $1.5 < F < 3.5$ if the wind is
accelerated by radiation pressure on ions in dust-free gas
(see \citealt{MC95,LB02,rpc4}).
To solve for $r_{L}$ and $r_{C}$, we take $F=2.5 \pm 1.0$
and assume that the observed radial velocity $v_{rad,obs}$ equals
the terminal velocity $v_\infty$.
If the latter assumption is incorrect,
the true $r_L$ will be smaller, so we call the value we obtain
with that assumption $r_{L,max}$.
Given the minimum velocity determined for trough A above,
$v>10,000$~\kms,we find
$r_{L,max} = 78_{-50}^{+74}~R_{Sch}=0.02_{-0.01}^{+0.02}$~pc and
$r_{C} \leq 186 \pm 75~R_{Sch}=0.04\pm{0.02}$~pc,
where the uncertainties on the values of the radii correspond
to the values assumed for $F$.
The minimum velocity determined for trough B above was
$8,000$~\kms, which yields
$r_{L,max} = 175_{-115}^{+165}~R_{Sch}=0.04_{-0.02}^{+0.04}$~pc and
$r_{C} \leq 350 \pm 75~R_{Sch}= \pm{3}$~pc.

\subsubsection{Acceleration}

The above estimate of $r_{L,max}$ assumes that the gas producing
trough B has reached maximum velocity, which may or may not be correct.
Here we explore some implications if that assumption is incorrect.
At the small radii inferred above, the gas may still be accelerating.
We can make an order of magnitude estimate of the expected acceleration
using some simple assumptions. We stress that these assumptions are not
unique, only illustrative.

The radial velocity of a radiatively-accelerated wind is approximately
$v(r)=v_\infty (1-r_L/r)^{1.15}$ \citep{mc97}.
The acceleration of the wind is
\begin{equation} 
a(r) = \frac{dv}{dt} = v\frac{dv}{dr}
= 1.15\frac{v_\infty^2 r_L}{r^2}\left(1-\frac{r_L}{r}\right)^{1.30}.
\end{equation}

If we assume a terminal velocity of $v_\infty=60,000$~\kms\ for trough B,
because the observed velocity of trough A shows that \civ\ absorption can
be seen to that high a velocity, then $r_L=78_{-50}^{+75}$ $R_{Sch}$.
If we set trough B's observed velocity $v_{rad,obs}=40,000~{\textrm{\kms}}=v(r_C)$,
we find $r_C=3.4 r_L = 265_{-170}^{+255}$ $R_{Sch}$.
(Incidentally, that yields a transverse velocity for trough B
of $v_{trans}=7,200^{+3000}_{-2100}$~\kms,
consistent with the lower limit on the transverse motion velocity we
determined for a flow-tube in \S~\ref{transmotion}).
The expected acceleration at $r_C=265$ $R_{Sch}$ for a wind launched at
$r_L=78$ $R_{Sch}$ is $35_{-11}^{+17}$~\kms~day$^{-1}$
(the maximum acceleration in that model is 86~\kms~day$^{-1}$).

This value is much larger than previous measurements of accelerating
BAL winds. For example, \citet{HS07} measured an acceleration
in a \civ\ trough found in SDSS J024221.87$+$004912.6 at
approximately $0.1$~\kms~day$^{-1}$.
The acceleration in J0230, if confirmed, would be the largest ever
detected in a BAL outflow.

Using the Gemini South telescope, we have obtained a new spectral epoch of
J0230 roughly 100 rest-frame days after the GEM3 epoch of this work.
If the above transverse motion variability model is correct, then
we predict trough B's centroid velocity will have
increased in velocity by $3,500_{-2400}^{+5200}$~\kms\ in that data.
The results of the new observations will be presented in a future
paper (Rogerson et al., in preparation).

This analysis was not done for trough A because we have no reliable
terminal velocity to suggest the cloud might accelerate to.

\subsection{Pure Ionization Parameter Variability Model}\label{ion}

In this model, we assume the absorbing clouds are not moving across the
emitting region of the quasar, and thus any variability observed in troughs
A and B is due to changes in the ionization parameter of the absorbing
clouds. In Filiz Ak et al. (2012, 2013), the authors observed coordinated
variability of distinct \civ\ BAL troughs in the same quasar, even
if the troughs are separated by as much as 10,000$-$20,000~\kms. Other
studies, such as \citet{GH15}, observed BAL troughs to vary across
the entire trough, rather than distinct sections. We do not
observe either of these behaviours in J0230: we find no significant
evidence for coordinated variations between troughs A and B
(they are separated by $\sim$15,000~\kms), and
we observe distinct regions of the absorption profiles to vary,
while others do not (specifically in trough B, see
\S~\ref{coord}). Nevertheless,
if we assume the changes observed in the troughs are due to an
ionization state change, we can place constraints on the physical
properties of the absorbing gas. Note that in this model
only fully saturated troughs will not vary.

The two absorbers responsible for troughs A and B
cannot have the same distances and densities
(including density as a function of velocity) to explain
the two trough's different responses to the same underlying
ionizing flux. The exception would be if the absorber closer
to the quasar significantly reduces the ionizing flux reaching
the absorber farther away. Whether the effect is significant or
not depends on the optical depth to ionizing radiation of the
absorber closer to the quasar.

Below, we assume that faster-responding gas has higher density.
If the changes in trough A are due purely to ionization
parameter variability, then the high-velocity part of this trough
has higher density (it responded more quickly, and then vanished).
If the changes in trough B are due purely to ionization
parameter variability, then the low-velocity part of trough B
has higher density (it responds faster to ionizing flux changes),
and the density drops off with increasing velocity.

One possible pure ionization variability scenario is the following.
Prior to SDSS, the ionizing flux $F_{ion}$ was high, leading to weak absorption.
Between SDSS and BOSS1, $F_{ion}$ decreased,
leading to an increase in \civ\ absorption (dense trough A appears).
After BOSS2, $F_{ion}$ recovered somewhat, leading to weaker trough A absorption.
Between BOSS2 and GEM1, lower-density trough B appears in response to the earlier decrease in $F_{ion}$.
The above scenario suggests that, barring any major future ionizing
flux variability, both trough A and trough B will decrease in strength
with time. Any other trough that appears will show slower evolution
in its EW than trough B does, due to the new trough's required lower density.





\subsubsection{Ionization constraints on electron density and distance}
\label{ionconstraint}

Constraints can be placed on the distance from the continuum source
to the absorbing gas, as well as the density of that gas
using the timescale of the variability in the absorption.
This approach has been used in multiple works
(see \citealt{HB95}, \cite{HB97}, \cite{NH04},
\citealt{AE12}, and references therein).
Below, we reproduce the approach taken in \citet{GH15}.

Consider gas initially in photoionization equilibrium
in the case where the ionization rate out of
ionization stage $i$ changes from its equilibrium value
$I_i$ to $(1+f)I_i$,
and the rate out of stage $i-1$ changes from $I_{i-1}$ to
$(1+f)I_{i-1}$,\footnote{Where we have assumed the fractional
change for $I_{i}$ and $I_{i-1}$ is the same.}
where $f$ is the fractional change in $I_i$.
Immediately after this change:
\begin{multline}
\frac{dn_i}{dt} = -f n_i I_i + f n_{i-1} I_{i-1} \\
+ [-n_i(I_i+R_{i-1})+n_{i-1}I_{i-1}+n_{i+1}R_i = 0]
\end{multline}
where the quantity in brackets is the equilibrium value of
$\frac{dn_i}{dt}$ and is therefore zero.
In equilibrium, $n_{i+1}/n_i = I_i/R_i$ 
where $R_i$ is the recombination rate to stage $i$,
because appearance/increase of stage $i$ by recombination from stage $i+1$
must be balanced by appearance/increase of stage $i+1$ by ionization from
stage $i$.
Thus we can substitute $n_{i-1}I_{i-1}=n_iR_{i-1}=n_i\alpha_{i-1} n_e$
(using $R_{i-1}=\alpha_{i-1} n_e$, where $\alpha_{i-1}$ is the recombination
coefficient to stage $i-1$) and rewrite $\frac{dn_i}{dt}$ as
\begin{equation}
\frac{dn_i}{dt} = -f n_i I_i + f n_i \alpha_{i-1} n_e
\end{equation}
which can be written as
\begin{equation}
\frac{dn_i}{n_i} \equiv \frac{dt}{t_i^*} ~{\rm with}~ t_i^* =
\left[ -f \left(I_i-n_e\alpha_{i-1}\right) \right]^{-1}
\end{equation}
which is an equation for variations on a characteristic
timescale $t_i^*$: $n_i(t) = n_i(0)\exp(t/t_i^*)$.

To summarize, for gas which is initially in photoionization equilibrium,
the characteristic timescale for density changes in ionization stage $i$
of some element in response to an ionizing flux change can be written as
$t_i^*$ above (a modified version of Eq.~10 of \citealt{AE12}),
where $-1<f<+\infty$ is the fractional change in
$I_i$, the ionization rate per ion of stage $i$ [$I_i(t>0) = (1+f)I_i(t=0)$],
$\alpha_{i-1}$ is the recombination coefficient to ionization stage $i-1$
of the ion, and a negative timescale represents a decrease in $n_i$ with time.
Note that this equation only considers photoionization
processes; collisional processes are neglected.
Gas which shows varying ionic column densities is not in a steady state by
definition, but such gas can still be in equilibrium with a varying ionizing
flux if its $t_i^*$ is considerably shorter than the flux variability
timescale (\S~6 of \citealt{KK95}).
For optically thin gas at distance $r$ from a quasar with luminosity
$L_\nu$ at frequency $\nu$, the ionization rate per ion of stage $i$
is given by
\begin{equation}
I_i=\int_{\nu_i}^{\infty}\frac{(L_\nu/h\nu)\sigma_\nu}{4{\rm \pi} r^2} d\nu
\end{equation}
where $\sigma_\nu$ is the ionization cross-section for photons of energy $h\nu$.

If the absorbing gas is far enough from the quasar that
$I_i\ll n_e\alpha_{i-1}$, then the relevant timescale
is $t_{rec}=1/f n_e\alpha_{i-1}$ (which is just the recombination time
of the ion in the $f=-1$ case where the ionizing flux drops to zero),
and the observed absorption variability timescale constrains the density
of the absorber.
However, if the absorbing gas is close enough to the quasar that
$I_i\gg n_e\alpha_{i-1}$, then the relevant timescale is $t_i=-1/f I_i$
and the absorption variations of the ion reflect the ionizing flux variations
of the quasar, with no density constraint derivable just from absorption
variations.\footnote{
No constraint on $n_e$ is derivable even though we can write the timescale as
\begin{equation}
t_i^* = \left[ -f \alpha_i n_e \left(\frac{n_{i+1}}{n_i} - \frac{\alpha_{i-1}}{\alpha_i}\right) \right]^{-1}
\end{equation}
(see equation (2) in \citealt{HB97}, equation (3) in \citealt{AC15})
because in our case $n_{i+1}/n_i = n_{\textrm{\cv}}/n_{\textrm{\civ}}$, and that
ratio increases more rapidly than $n_e$ decreases as the
ionization parameter increases \citep{KM82}.}
An observed timescale for variations in optically thin absorption therefore
constrains the absorbing gas to either have a density $n_e>n_{min}$ and
$r>r_{equal}$, where $r_{equal}$ is the distance at which $I_i=n_{min}\alpha_{i-1}$,
or to be located at $r<r_{equal}$ with almost no constraint on the density.

As noted in \citealt{AE12}, there are limitations to using timescale
arguments to infer physical characteristics of an absorber.
In that work, the authors
indicate "a more physically motivated approach is to use lightcurve
simulations that are anchored in our knowledge of the power spectrum
behaviour of observed AGN lightcurves;" however, such detailed
work is not justified by the relatively scarce data available for J0230.

To determine the constraints on the emergence of troughs A and B,
we assume a temperature of
$\log T = 4.3$
(\citealt{K99}) 
so that the recombination coefficient is
$\alpha_{C\,III}=2.45\times 10^{-11}$ cm$^3$ s$^{-1}$
(from the CHIANTI online database; see \citealt{DL97}, \citealt{LY13}).
For the simple case of the ionizing flux dropping to zero,
$f=-1$ and the timescale $t^*_i$ can be approximated as the
recombination time, $t_{rec} \sim 1/n_e\alpha_{C\,III}$.

Using the time between observations of SDSS and BOSS1 for trough A
of 652 days as an upper limit to the recombination time, we calculate
a lower limit on the density of the gas to be $n_{e,A} \geq 724$ cm$^{-3}$.
Using the lower limit density of $n_{e,A} \geq 724$ cm$^{-3}$, we calculate
the minimum distance from the quasar at which that lower limit is valid.
From its observed flux density at rest-frame 3000~\AA, our quasar has $L_{bol}=8.45\times 10^{46}$ erg s$^{-1}$.
We adopt the spectral energy distribution of \citet{DB10}
to calculate $L_\nu$. Therefore, if the emergence of trough A is due
to ionization variability, the absorber either has a density of
$n_{e,A} \geq 724$ cm$^{-3}$ and is at $r_{equal,A} \ge 2.00 {\rm ~kpc}$, or is
at $r < 2.00 {\rm ~kpc}$ with no constraint on the density.

Trough B emerged between BOSS2 and GEM1; a period of 307 days.
Using this as an upper limit to the
recombination time, we perform the same calculation and determine
if the appearance of trough B is due to
ionization variability, the absorber either has a
density $n_{e,B} \geq 1540~{\rm cm}^{-3}$ and is at
$r_{equal,B} \ge 1.37 {\rm ~kpc}$, or is $r < 1.37 {\rm ~kpc}$
with no constraint on the density.

Our values for $n_e$ are one or two orders of magnitude lower
than those found in \citet{GH15} and \citet{CH13}
(which found values
$\sim10^5$ cm$^{-3}$ and our values of $r_{equal}$ are
10 times larger those works (which found values $\sim$100 pc).
Further, our values of $r_{equal}$ are much higher than the
launching radius of BAL winds expected from theoretical
work, which predict $\sim10^{-3}$~pc (e.g., \citealt{MC95}).
Nonetheless, 
other works have reported outflow radii on similar scales to
that we infer for J0230 in a pure ionization variability model
(see Table~10 of \citealt{DB10}, and references therein),
and the radius at which a BAL wind is observed is not
necessarily the radius at which the wind is launched
(e.g., \citealt{FQ12}).

Finally, we can place an upper limit constraint on $n_e$ by
searching for absorption features from other ions of carbon,
specifically \cii\ \lam1335~\AA. Given the minimum density and
$r_{equal}$ distances determined above, troughs A and B are created by
absorbers with an ionization parameter of
$U_H \simeq 0.06$.\footnote{This is for our assumed SED
from \citealt{DB10},
$U_H = Q_H/4{\rm \pi} n_H c$, with $Q_H = 6.08\times10^{56}$
hydrogen-ionizing photons s$^{-1}$ and $n_H = 0.82n_e$.}
If we lowered the
ionization parameter by a factor of $\sim$50, either by gas at
larger radii or at higher density, the resulting
ionization state would yield \cii\ absorption roughly half
as deep as the observed \civ\ (see Figure~3 of \citealt{HB95}).

A reduction by a factor of $\sim$50 in ionization parameter gives us
upper limits to both the minimum density and the $r_{equal}$;
Therefore, the absorber that caused the emergence of trough A
has $724~{\rm cm}^{-3} \le n_{e,A} \le 3.62\times 10^4(r_{equal,A}/r)^2~{\rm cm}^{-3}$ and is between $r_{equal,A} \le r \le 7r_{equal,A}$.
Similarly, the absorber that caused the emergence of trough B
is constrained by $1540~{\rm cm}^{-3} \le n_{e,B} \le 7.70\times 10^4(r_{equal,B}/r)^2~{\rm cm}^{-3}$ and $r_{equal,B} \le r \le 7r_{equal,B}$.
These upper limits only work in the scenario where we approximate
the recombination time as $t_{rec} \sim 1/n_e\alpha_{C\,III}$.

\begin{figure}
\centering
\includegraphics[width=\columnwidth]{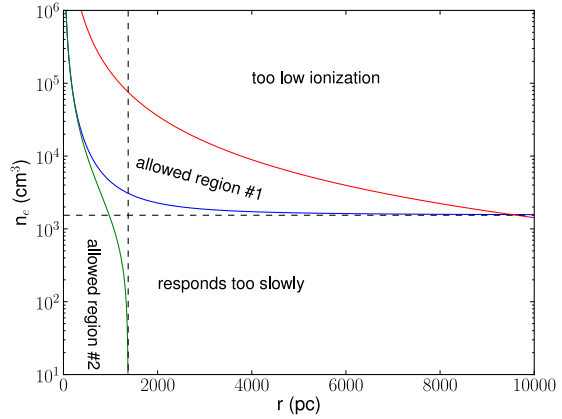}
\caption{The possible combinations of density and distance for the
gas that created trough B. The horizontal dashed line represents the density if
the ionizing flux dropped to zero and we use the time between observations
as the recombination time. The vertical dashed line represents minimum
distance from the quasar at which the lower limit to the density is
valid.}
\label{denreq}
\end{figure}

In Figure~\ref{denreq}, we have plotted the possible values of the
density of the absorbing gas $n_e$ and the distance the absorber
is from the source $r$, given constraints imposed by the timescale
arguments above for trough B. The vertical and horizontal dashed lines
are the locations of the $r_{equal,B}$ and the minimum electron
density $n_{e,B}$, respectively.
Any combination of parameters above the red line would have too high
a density or too far a distance to be ionized to \civ\ (and lead
to the upper limit arguments above). There
is also a region of too low density or too far away that
requires too long a timescale for the proper response. The allowed
regions 1 (between the red and blue curves) and 2 (to the left of
the green curve) represent the combinations of parameters possible.
There is also a region of too high ionization at low densities and
small radii which is not visible at the scale shown. Note that in
the discussion at the end of \S~\ref{ion} we assumed that
faster-responding gas has higher density, although from
Figure~\ref{denreq} that is only certain if the gas is at $r>r_{equal}$.
A corresponding plot for trough A would look similar.






\section{Summary}

We have presented the discovery and analysis of two extremely
high-velocity and highly-variable \civ\ BAL troughs in the quasar
SDSS~J023011.28+005913.6. We retrieved 4 spectra of J0230 from
the SDSS$+$BOSS archives, and obtained 3 of our own spectra
using the Gemini Observatory. The longest time between
observations was $\sim650$ rest-frame days, and the shortest
was $\sim10$ rest-frame days.

\begin{enumerate}

\item We discovered a \civ\ BAL trough outflowing from J0230
at $\sim$60,000~\kms\ (trough A), the largest velocity of a
BAL wind observed to-date. During follow up observations, we
discovered a second \civ\ BAL trough outflowing at
$\sim$40,000~\kms\ (trough B). See Figure~\ref{all}.

\item In troughs A and B we observed variability of
both the depth and shape of the troughs on scales as
short as 10 days in the rest-frame. See Table~\ref{measure}.

\item A dataset of six spectral epochs straddling the
emergence of both troughs allowed us to rule out some
simple models of bulk motion as the origin of the variability.
It also allowed us to propose and test more complex and
realistic models of bulk motion, such as flow-tube geometries
and an augmented `crossing disks'+SS73 scenario.
See \S~\ref{transmotion}.

\item We found the variability of trough A
is best explained by a `crossing disk' traversing a SS73 disk
at velocities between $10,000 < v~(\textrm{\kms}) < 18,000$.
This model and velocity range allow us to predict trough A
will disappear between $350 < t~(\textrm{days}) < 1,500$ after
our last observation (GEM3). See \S~\ref{transmotion}.

\item Trough B is best explained by a flow-tube that has recently
moved into the line of sight, travelling in the velocity range
$8,000 < v~(\textrm{\kms}) < 56,000$. In this scenario, we have
no constraint on how far a flow-tube extends, and thus cannot
predict when trough B will disappear. See \S~\ref{transmotion}.

\item Given some simple, conservative assumptions in a
transverse velocity model, we constrained the distance
from the black hole to the absorbing gas responsible for trough A
$r_{C} \leq 186 \pm 75~R_{Sch}=0.04\pm{0.02}$~pc given $v_{trans}>10,000$~\kms\
and for trough B we contrain the distance to be
$r_{C} \leq 350 \pm 140~R_{Sch}=0.07\pm{0.03}$~pc for $v_{trans}>8,000$~\kms.
See \S~\ref{constraindist}.

\item If we assume changes to the ionization parameter
is the reason for the variability observed, the absorber
responsible for trough A either has
$724~{\rm cm}^{-3} \le n_{e,A} \le 3.62\times 10^4(r_{equal,A}/r)^2~{\rm cm}^{-3}$ and is between $r_{equal,A} \le r \le 7r_{equal,A}$,
or is at $r < 2.00 {\rm~kpc}$ with no constraint on the density.
Similarly, the absorber that caused the emergence of trough B
is either constrained by $1540~{\rm cm}^{-3} \le n_{e,B} \le 7.70\times 10^4(r_{equal,B}/r)^2~{\rm cm}^{-3}$ and $r_{equal,B} \le r \le 7r_{equal,B}$, or is at $r_{equal} \ge 1.37 {\rm~kpc}$, or is at $r < 1.37 {\rm~kpc}$ with no constraint on the density. See \S~\ref{ionconstraint}.

\end{enumerate}

Given the results above, we cannot rule out bulk motion or
ionization changes as models of BAL variability. More observations
of J0230 will, however, allow us to test if our predictions of
how the troughs will vary in the future are accurate.

\section{Acknowledgments}
We thank the referee for a thorough and helpful review.
PBH and JAR are supported by NSERC.
NFA would like to acknowledge financial support from TUBITAK (115F037).
WNB would like to thank NSF grant AST-1516784.

Based on observations for
Program IDs GN-2013B-Q-59, GN-2013B-Q-39, and GS-2013B-Q-21
obtained at the Gemini Observatory (processed using the Gemini IRAF package),
which is operated by the
Association of Universities for Research in Astronomy, Inc., under a cooperative agreement
with the NSF on behalf of the Gemini partnership: the National Science Foundation
(United States), the National Research Council (Canada), CONICYT (Chile), the Australian
Research Council (Australia), Minist\'{e}rio da Ci\^{e}ncia, Tecnologia e Inova\c{c}\~{a}o
(Brazil) and Ministerio de Ciencia, Tecnolog\'{i}a e Innovaci\'{o}n Productiva (Argentina).

The authors wish to recognize and acknowledge the very significant cultural role and reverence that the summit of Mauna Kea has always had within the indigenous Hawaiian community. We are most fortunate to have the opportunity to conduct observations from this mountain.

Funding for SDSS-III has been provided by the Alfred P. Sloan Foundation, the Participating Institutions, the National Science Foundation, and the U.S. Department of Energy Office of Science. The SDSS-III web site is http://www.sdss3.org/.

SDSS-III is managed by the Astrophysical Research Consortium for the Participating Institutions of the SDSS-III Collaboration including the University of Arizona, the Brazilian Participation Group, Brookhaven National Laboratory, Carnegie Mellon University, University of Florida, the French Participation Group, the German Participation Group, Harvard University, the Instituto de Astrofisica de Canarias, the Michigan State/Notre Dame/JINA Participation Group, Johns Hopkins University, Lawrence Berkeley National Laboratory, Max Planck Institute for Astrophysics, Max Planck Institute for Extraterrestrial Physics, New Mexico State University, New York University, Ohio State University, Pennsylvania State University, University of Portsmouth, Princeton University, the Spanish Participation Group, University of Tokyo, University of Utah, Vanderbilt University, University of Virginia, University of Washington, and Yale University.


\end{document}